\documentclass[11pt]{article}
\usepackage{moriondmod,epsfig}

\newcommand{\beq}{\begin{equation}}
\newcommand{\eeq}{\end{equation}}
\newcommand{\bea}{\begin{eqnarray}}
\newcommand{\eea}{\end{eqnarray}}

\newcommand{\gsim}{\lower.7ex\hbox{$
\;\stackrel{\textstyle>}{\sim}\;$}}
\newcommand{\lsim}{\lower.7ex\hbox{$
\;\stackrel{\textstyle<}{\sim}\;$}}

\def\lsim{\mathrel{\rlap{\lower3pt\hbox{\hskip0pt$\sim$}}
    \raise1pt\hbox{$<$}}}         
\def\gsim{\mathrel{\rlap{\lower4pt\hbox{\hskip1pt$\sim$}}
    \raise1pt\hbox{$>$}}}         

\newcommand{\bibit}[1]{\bibitem{#1}}

\newcommand{\aver}[1]{\langle #1\rangle}

\newcommand{\La}{\overline{\Lambda}}
\newcommand{\Si}{\overline{\Sigma}}
\newcommand{\Lam}{\Lambda_{\rm QCD}}

\newcommand{\tto}{\!\to\!}

\newcommand{\GeV}{\,\mbox{GeV}}
\newcommand{\MeV}{\,\mbox{MeV}}
\newcommand{\matel}[3]{\langle #1|#2|#3\rangle}
\newcommand{\state}[1]{|#1\rangle}

\begin{document}
\vspace*{4cm}
\title{QCD IN BEAUTY DECAYS: ~SUCCESSES AND PUZZLES}

\author{ N.~URALTSEV }

\address{INFN, Sezione di Milano,  Milano, Italy $^\dagger$
\footnotetext{$^\dagger$ On leave of
absence from Department of Physics, University of Notre Dame, 
Notre Dame, IN 46556, USA,
\\
and from Petersburg Nuclear Physics Institute, Gatchina, 188300 Russia. }}

\maketitle\abstracts{The status of the heavy quark expansion for
inclusive $B$ decays is briefly reviewed from the perspective of
confronting theory with data and of extracting the heavy quark
parameters. A good agreement between properly applied
theory and new precision data is observed. Some recent applications to the
exclusive heavy flavor transitions are addressed. I recall the
`$\frac{1}{2}\!>\!\frac{3}{2}$' paradox for the transitions into the
charm $P$-wave states.}

\vspace*{-282pt}\begin{flushright}
Bicocca-FT-04-6\\
UND-HEP-04-BIG\hspace*{.08em}04\vspace*{254pt}\\
\end{flushright}

Quantum Chromodynamics as the fundamental theory of strong forces has
vast applications at energy scales separated by many orders of
magnitude. One of its important uses is in heavy quark physics, in
particular electroweak decays of beauty particles.  Theory makes most
interesting dynamic statements about sufficiently inclusive heavy
flavor decays which admit the local operator product expansion
(OPE). These applications are important in the phenomenological
aspects -- they allow model-independent extractions of the underlying
CKM mixing angles $|V_{cb}|$ and $|V_{ub}|$ with record accuracy from
the data; likewise for fundamental parameters like $m_b$ and $m_c$. At
the same time studying inclusive decay distributions yields unique
information about QCD itself in the nonperturbative regime.

A few year old review of the principal elements of the heavy quark
theory can be found in Ref.~\cite{ioffe}; I closely follow its
nomenclature here. The field is still being actively developed, and
new connections with other problems of high-energy QCD may be at the
horizon.

Heavy quark theory derived informative dynamic consequences for a
number of exclusive transitions as well. Recent years finally
witnessed a more united approach to inclusive and exclusive decays
which previously have been largely separated. I will illustrate some
of the nontrivial connections of this sort in the last part of my
talk. 

Most elements of the heavy quark expansion for inclusive decays have
been elaborated in the 1990s together with applications to
extraction of the heavy quark parameters. They were not in the focus,
however. Moreover, there has been a belief that the theory predictions
are not in a good agreement with the available data, a feeling,
probably carried on into these days.

The last years were a turning point in this
respect. Experiments, including those at the now operating $B$ factories have
accumulated data sets of qualitatively different statistics and
precision, often of a higher `theory consumer value' as well. Certain
developments in theory match the progress. A more robust 
approach to the analysis has been put forward \cite{amst} and
applied in practice \cite{delphi}, and made more systematic \cite{slcm}. 
The perturbative
corrections for all inclusive semileptonic characteristics were
finally calculated \cite{slsf,trott}.

It is now the right time to critically review the theory standing when
confronted with the \newpage \noindent data.
It turns out that the formerly 
alleged problems are replaced by impressive agreement, and theory
often seems to work even better than can realistically be expected,
when pushed to the hard extremes.

\section{Inclusive semileptonic decays: theory vs.\ data}

The central theoretical result for the inclusive decay rates of heavy
quarks is that they are not affected by nonperturbative physics at the
level of $\Lam/m_Q$ (even though hadron masses, and, hence the phase
space itself, are), and the corrections are given by the local heavy
quark expectation values -- $\mu_\pi^2$ and $\mu_G^2$ to order
$1/m_Q^2$, etc.\ \cite{buvbs}. Today's status has quite 
advanced and allows to aim at an 1\% accuracy in $|V_{cb}|$ extracted from
$\Gamma_{\rm sl}(B)$. A similar approach to $|V_{ub}|$ is more
involved since theory has to conform with the practical necessity to
implement significant cuts to reliably reject the $b\tto c\,\ell\nu$
decays. Yet the corresponding studies are underway and a 5\% accuracy
seems realistic.

There are many aspects theory must address to target this level of
precision. One facet is perturbative corrections, a subject of
controversial statements for a long time. The reason goes back to
rather subtle aspects of the OPE. It may be partially elucidated by
Figs.~1 showing the relative weight of gluons with different momenta
$Q$ affecting the total decay rate and the average hadronic recoil
mass squared $\aver{M_X^2}$, respectively. The contributions in the
conventional `pole'-type perturbative approach have long tales
extending to very small gluon momenta below $500\MeV$, especially for
$\aver{M_X^2}$; the QCD coupling $\alpha_s(Q)$ grows uncontrollably
there. These would be a disaster for precision calculations manifest,
for instance, through a numerical havoc once higher-order corrections
are incorporated. Yet applying literally the Wilsonian prescription
for the OPE with explicit separation of scales in all strong
interaction effects, including the perturbative contributions,
effectively cuts out the infrared pieces! Not only the higher-order
terms emerge suppressed, even the leading-order corrections become
small and more stable. This approach applied to heavy quarks long ago
\cite{upsetblmopefive} implies that the precisely defined running
heavy quark parameters $m_{b}(\mu)$, $\La(\mu)$,
$\mu_\pi^2(\mu)$, ... appear in the expansion, rather than ill-defined
parameters like pole masses, $\La$, $-\lambda_1$ usually employed by
HQET. Then it makes full sense to extract these well-defined QCD
objects with high precision.

\begin{figure}[hhh]
\vspace*{-1.5pt}
\begin{center}
\mbox{\psfig{file=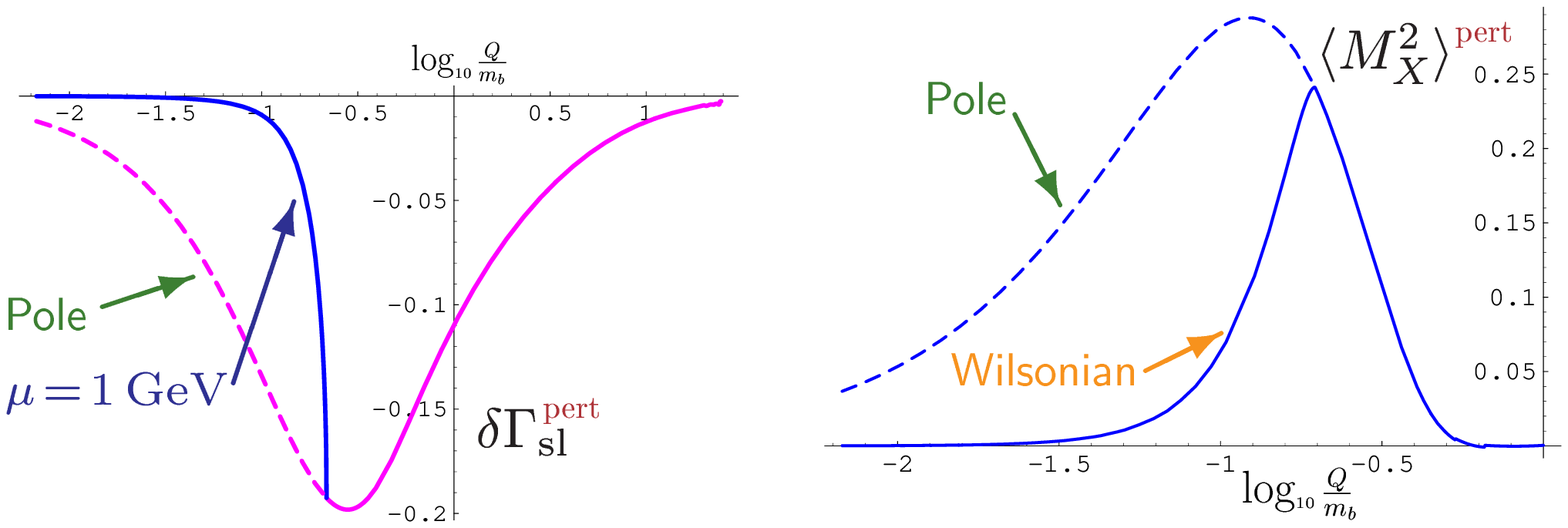,width=400pt}}\vspace*{-4pt}
\end{center}
\caption{The distribution over the perturbative gluon momentum 
in $\Gamma_{\rm sl}$ and in $\aver{M_X^2}$, 
for $b\tto c\,\ell \nu$.\vspace*{-1pt}} 
\end{figure}

The most notable of all the alleged problems for the OPE in the
semileptonic decays was, apparently, the
dependence of the final state invariant hadron mass on the lower cut
$E_{\rm cut}^\ell$ in the lepton
energy: theory seemed to fall far off \,\cite{ligman} of the
experimental data. The robust approach, on the contrary 
appears to describe it well \cite{misuse}, as illustrated by
Figs.~2. The second moment of the same distribution also
seems to perfectly fit theoretical expectations \cite{slcm,slsf} with
the heavy quark parameters extracted by BaBar from their data
\cite{babarprl}. 

\begin{figure}[hhh]
\vspace*{3pt}
\begin{center}
\hspace*{-10pt}
\mbox{\psfig{file=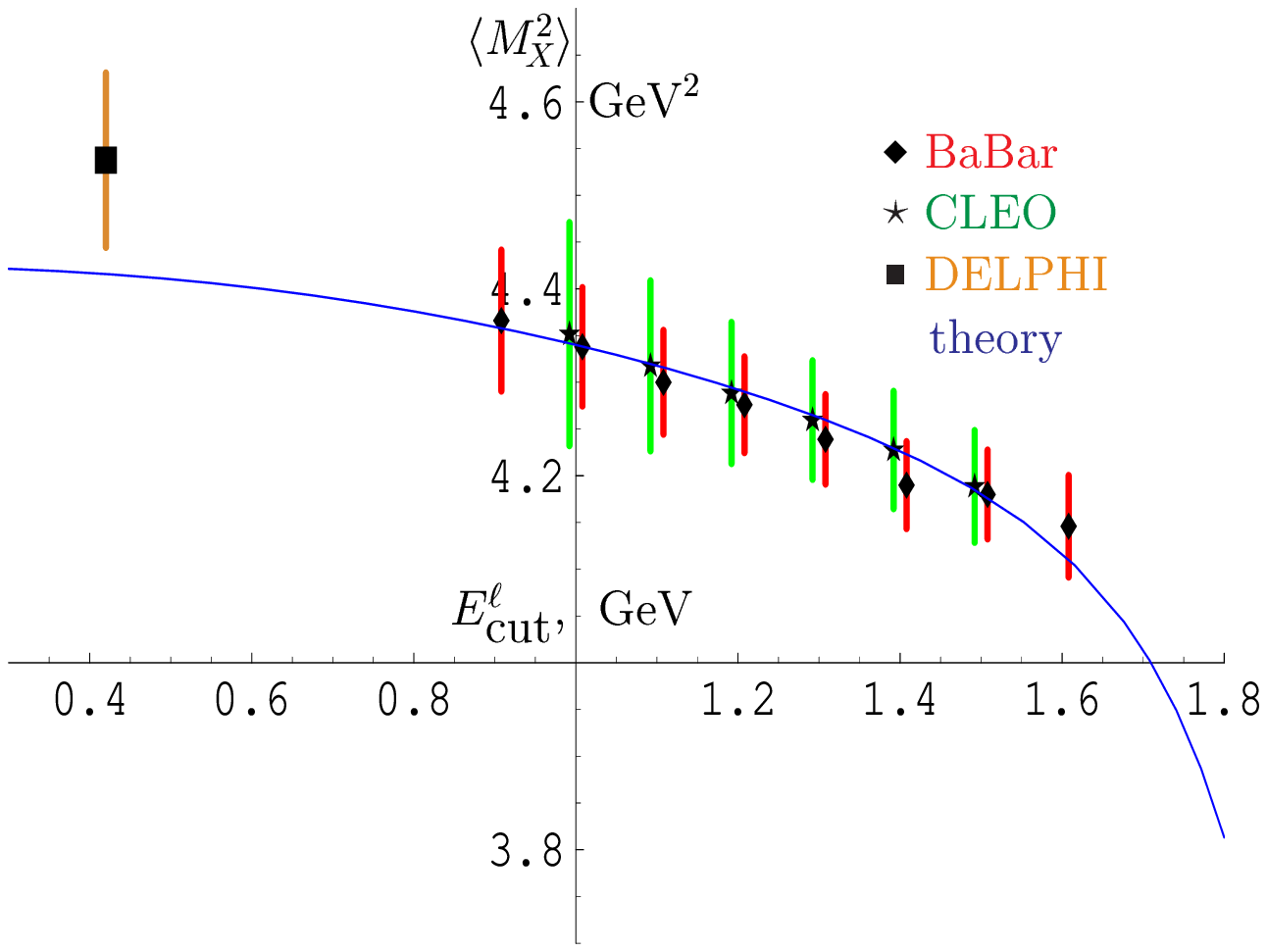,width=200pt}}\hspace*{40pt} 
\mbox{\psfig{file=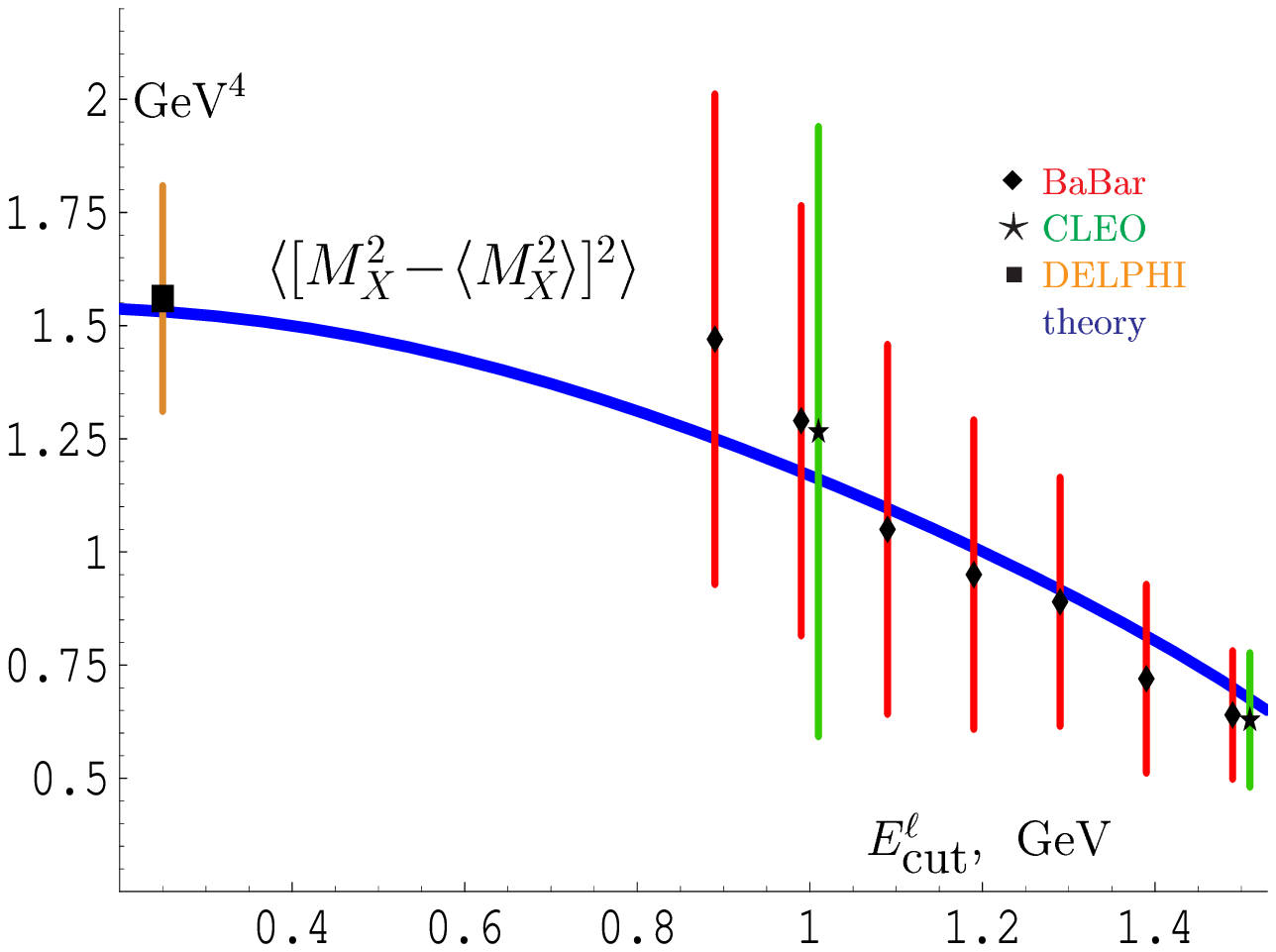,width=200pt}}\vspace*{-4pt}
\end{center}
\caption{Hadron mass moments dependence on the lepton energy cut. 
\vspace*{-1pt}}
\end{figure}

The problem in the calculations of Ref.~\cite{ligman} has not been
traced in detail; its approach had a number of vulnerable
elements. I suspect an algebraic mistake in the perturbative
corrections there. On the other hand, the principal perturbative
scheme in Ref.~\cite{ligman} is based on what the authors call
``$\Upsilon$-expansion'' for the perturbative series. That procedure
assumes a rather ad hoc reshuffling among different orders in
$\alpha_s$, and its theoretical validity is questionable.

Another point of possible data vs.\ theory discrepancy used to be an 
inconsistency between the values of the heavy quark parameters
extracted from the semileptonic decays and from 
the photon energy moments \cite{ligbsg} in
$B\tto X_s\!+\!\gamma$. It has been pointed out, however \cite{misuse}
that with relatively high experimental cuts on $E_\gamma$ the actual
`hardness' ${\cal Q}$ significantly degrades compared to $m_b$, thus
introducing the new energy scale with ${\cal Q}\simeq 1.2\GeV$ at
$E_{\rm cut}^{\gamma}= 2\GeV$. Then the terms exponential in ${\cal
Q}$ left out by the conventional OPE, while immaterial under normal
circumstances, become too important. This is illustrated by Figs.~3
showing the related `biases' in the extracted values of $m_b$ and
$\mu_\pi^2$. Accounting for these effects appeared to turn
discrepancies into a surprisingly good agreement between all the
measurements \cite{misuse}.

\begin{figure}[hhh]
\vspace*{-1pt}
\begin{center}
\mbox{\psfig{file=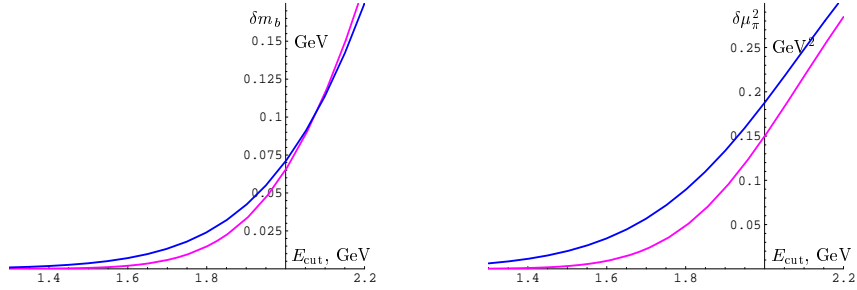,width=320pt}}\vspace*{-18pt} 
\end{center}
\caption{`Exponential' biases in $m_b$ and $\mu_\pi^2$ due to
the lower cut on photon energy in $B\tto X_s+\gamma$.\vspace*{-3pt}} 
\end{figure}

The comparison of data with the OPE-based theory is examined in more
detail in the talk by H.~Flaecher (BaBar), and I refer to it for further
intriguing observations \cite{flaecher}. A similar analysis of the new
precision CLEO data \cite{shepherd} is underway. 

BELLE recently reported photon energy moments down to the energy
$1.8\GeV$ \cite{belle}, which suppresses the impact of biases:
\beq
\aver{E_\gamma}\!=\!2.289\pm 0.026_{\rm stat}\pm 0.0034_{\rm sys} \GeV,\;\; 
\aver{(E_\gamma \!-\!\aver{E_\gamma})^2}\!=\!0.0311
\pm 0.0073_{\rm stat}\pm 0.0063_{\rm sys} \GeV^2.
\label{40}
\eeq
The theoretical expectations based on the central BaBar values of the
parameters with $m_b\!=\!4.611\GeV$,  $\mu_\pi^2\!=\!0.447\GeV^2$, for the
moments with $E_{\rm cut}^{\gamma}\!=\!1.8\GeV$
are 
\beq
\aver{E_\gamma} \simeq 2.317 \GeV\;,\qquad  
\aver{(E_\gamma -\aver{E_\gamma})^2}\simeq 0.0329 \GeV^2\;,
\label{42}
\eeq
again in a good agreement. 

As a brief summary, data show a good agreement with the properly
applied heavy quark theory. In particular, it appears that\\
$\bullet$ Many underlying heavy quark parameters have been accurately
determined directly from experiment.\\
$\bullet$ Extracting $|V_{cb}|$ from $\Gamma_{\rm sl}(B)$ has
high accuracy and rests on solid grounds.\\
$\bullet$ We have precision checks of the OPE-based theory at the
level where nonperturbative effects play significant role.

In my opinion, the most nontrivial and critical for theory is the 
consistency found between hadronic mass and lepton energy moments, in
particular $\aver{M_X^2}$ vs.\ $\aver{E_\ell}$.\footnote{I would not
rate high in this respect the success itself in describing the
dependence of the ($M_X^2$) moments on $E_{\rm cut}$ in the wide
interval, due to expected decrease in theory predictability above
$E_{\rm cut}\!\simeq\! 1.35\GeV$, see Ref.~\cite{fpcp03}; 
different opinions exit here, though.} This is a sensitive check of the
nonperturbative sum rule for $M_B\!-\!m_b$, at the precision 
level 
of higher power corrections. It is interesting to note in this respect
that a particular combination of the quark masses, $m_b\!-\!0.74m_c$
has been determined in the BaBar analysis with only $17\MeV$ error bar!
This illustrates how $|V_{cb}|$ can be obtained with the high
precision: the semileptonic decay rate $\Gamma_{\rm sl}(B)$ is driven
by nearly the same combination \cite{imprec}.
\vspace*{2mm}

\section{ A `BPS' expansion}

The heavy quark 
parameters as they emerge from the BaBar fit are 
close to the theoretically expected values,  $m_b(1\GeV)\!\simeq\!
4.60\GeV$, $\mu_\pi^2(1\GeV)\!\simeq\!
0.45\GeV^2$, $\rho_D^3(1\GeV)\!\simeq\! 0.2\GeV^3$. The precise value, in
particular of $\mu_\pi^2$, is of considerable theoretical interest. It
is essentially limited from below by the known chromomagnetic
expectation value \cite{ineq}:
\beq
\mu_\pi^2(\mu)> \mu_G^2(\mu), \qquad \mu_G^2(1\GeV)\simeq 
0.35^{+.03}_{-.02}\GeV^2,
\label{60}
\eeq
and experiment seem to suggest that this bound is not too far from
saturation. This is a peculiar regime where the so-called heavy
quark sum rules \cite{ioffe}, the exact relations 
for the transition amplitudes
between very heavy flavor hadrons, become highly 
constraining. 

One consequence of the heavy quark sum rules is the lower bound 
\cite{newsr} on
the slope of the IW function $\varrho^2\!>\!\frac{3}{4}$. There are 
also upper bounds which turn out quite
restrictive once $\mu_\pi^2$ is close to $\mu_G^2$, \,say
\beq
\varrho^2 \!- \mbox{$\frac{3}{4}$} \lsim  0.3 \;\; \mbox{ ~if~ }\;\; 
\mu_\pi^2(1\GeV)\!-\! \mu_G^2(1\GeV) \lsim 0.1\GeV^2.
\label{64}
\eeq
This illustrates the magic power of the comprehensive 
heavy quark expansion in QCD: the
moments of the inclusive semileptonic decay distributions can tell 
us, for instance, about the formfactor for $B\tto D$ or $B\tto D^*$ decays.

Another application is the $B\tto D\,\ell \nu$ amplitude near zero
recoil. Expanding in $\mu_\pi^2\!-\!\mu_G^2$ an accurate estimate was
obtained \cite{bps}
\beq
\frac{2\sqrt{M_B M_D}}{M_B+M_D} f_+(0)\simeq 1.04\pm 0.01\pm 0.01\;.
\label{70}
\eeq
In fact, $\mu_\pi^2\simeq \mu_G^2$ is a remarkable point for
physics of $B$ and $D$ mesons, since the equality implies a functional
relation $\vec\sigma_b\vec\pi_b\state{B}\!=\!0$. Some of the Heavy
Flavor symmetry relations (but not those following from the spin
symmetry) are then preserved to {\sf all orders} in $1/m_Q$. This
realization led to a `BPS' expansion \cite{chrom,bps} where
the usual heavy quark expansion was combined with expanding around
the `BPS' limit $\vec\sigma_b\vec\pi_b\state{B}\!=\!0$. 

There are a number of miracles in the `BPS' regime. They  
include $\varrho^2\!=\!\frac{3}{4}$ and 
$\:\rho_{LS}^3\!=\!-\rho_D^3$; a complete discussion can be found in
Ref.~\cite{bps}. 
Some intriguing ones are:\\
$\bullet$ No power corrections to the relation
$M_P\!=\!m_Q+\La$ and, therefore to $m_b\!-\!m_c=M_B\!-\!M_D$. \\
$\bullet$ For the $\;B\!\to\! D\;$ amplitude the heavy quark limit 
relation between
the two formfactors 
\beq
f_-(q^2)=-\frac{M_B\!-\!M_D}{M_B+M_D}\; f_+(q^2)
\label{106}
\eeq
does not receive power corrections.\\
$\bullet$ For the zero-recoil $\;B\!\to\! D\;$ amplitude all
$\,\delta_{1/m^k}\,$ terms vanish.\\
$\bullet$ For the zero-recoil formfactor $\,f_+\,$ controlling decays 
with massless leptons
\beq
f_+((M_B\!-\!M_D)^2)=\frac{M_B+M_D}{2\sqrt{M_B M_D}}
\label{108}
\eeq
holds to all orders in $1/m_Q$.\\
$\bullet$ At arbitrary velocity power corrections in $\;B\!\to\! D\;$
vanish,
\beq
f_+(q^2)=\frac{M_B+M_D}{2\sqrt{M_B M_D}} \;\,\mbox{{\large$ 
\xi$}}\!\left(\mbox{$\frac{M_B^2+M_D^2-
\raisebox{.6pt}{\mbox{{\normalsize $q^2$}}}}{2M_BM_D}$}\right)
\label{110}
\eeq
so that the $\;B\!\to\! D\;$ decay rate directly yields 
Isgur-Wise function $\xi(w)$.

Since the `BPS' limit cannot be exact in actual QCD, we need to
understand the accuracy of its predictions.  The dimensionless
parameter $\beta$ describing the deviation from `BPS' is not tiny,
similar in size to the generic $\,1/m_{c}\,$ expansion parameter, and
relations violated to order $\beta$ may in practice be more of a
qualitative nature.  However, the expansion parameters like
$\mu_\pi^2\!-\!\mu_G^2 \propto \beta^2$ can be good enough. One can
actually count together powers of $1/m_c$ and $\beta$ to judge the
real quality of a particular heavy quark relation.  In fact, the
classification in powers of $\beta$ to {\tt all orders} in $1/m_Q$ is
possible.

Relations (\ref{106}) and (\ref{110}) for the $B\!\to\!D$
amplitudes at arbitrary velocity can get first order corrections in
$\beta$, and may be not very accurate.  Yet the slope $\varrho^2$ 
of the IW function differs from $\frac{3}{4}$ only at 
order $\beta^2$.
Some other important `BPS' relations hold up to order $\beta^2$:\\
$\bullet$ $M_B\!-\!M_D=m_b\!-\!m_c$ and $M_D=m_c\!+\!\La$ \\
$\bullet$ Zero recoil matrix element $\matel{D}{\bar{c}\gamma_0 b}{B}$
is unity up to ${\cal O}(\beta^2)$\\
$\bullet$ Experimentally measured $B\!\to\!D$ formfactor $f_+$ near
zero recoil receives only second-order corrections in $\beta$ to all
orders in $1/m_Q$:
\beq
f_+\left((M_B\!-\!M_D)^2\right) = \frac{M_B\!+\!M_D}{2\sqrt{M_BM_D}} \;\,
+ {\cal O}(\beta^2)\;.
\label{116}
\eeq
The latter is an analogue of the Ademollo-Gatto theorem for the `BPS'
expansion. 

As a practical application, Ref.~\cite{bps} derived a rather accurate
estimate for the formfactor $f_+(0)$ in the $B\tto D$ transitions,
Eq.~(\ref{70}), incorporating terms through $1/m_{c,b}^2$. The largest
correction, $+3\%$ comes from the short-distance perturbative
renormalization; power corrections are estimated to be only about $1\%$.
\vspace*{2mm}

\section{The {\boldmath `$\frac{1}{2}>\frac{3}{2}$'} puzzle}

So far I have discussed mostly the success story of the heavy quark
expansion for semileptonic $B$ decays. At the same time I feel
important to recall the so-called `$\frac{1}{2}>\frac{3}{2}$' puzzle
related to the question of saturation of the heavy quark sum
rules. Raised independently by two teams \cite{rev,ioffe,orsaysig}
including the heavy quark group in Orsay, it has been around for quite
some time, yet did not attract much attention so far.

There are two basic classes of the sum rules in the Small Velocity, 
or Shifman-Voloshin (SV) heavy quark limit. First are spin-singlet
which relate $\varrho^2$, $\La$, $\mu_\pi^2$, $\rho_D^3$,... to the
excitation energies $\epsilon$ and transition amplitudes squared 
$|\tau|^2$ for
the $P$-wave states. These sum rules get contributions from both
$\frac{1}{2}$ and $\frac{3}{2}$ $P$-wave states, i.e.\ those where 
the spin $j$ of the light cloud is $\frac{1}{2}$ or $\frac{3}{2}$. 

The second class are `spin' sum rules, they express similar relations
for $\varrho^2\!-\!\frac{3}{4}$, $\La\!-\!2\Si$,
$\mu_\pi^2\!-\!\mu_G^2$, etc.  These sum rules include only
$\frac{1}{2}$ states.

The spin sum rules strongly suggest that the
$\frac{3}{2}$ states dominate over $\frac{1}{2}$ states, having
larger transition amplitudes $\tau_{3/2}$. In fact, this automatically
happens in all quark models respecting Lorentz covariance and the
heavy quark limit of QCD; an example are the Bakamjian-Thomas--type 
quark models developed at Orsay \cite{orsayqm}. 

The lowest $\frac{3}{2}$ $P$-wave excitations of $D$ mesons, $D_1$ and
$D_2^*$ are narrow and well identified in the data. Their
contribution to the sum rules appears too small, however, with
$|\tau_{3/2}|^2\approx 0.15$ \cite{leib}. Wide  $\frac{1}{2}$ 
states denoted by 
$D_0^*$ and $D_1^*$ are more copiously produced; they can, in
principle, saturate the singlet sum rules. However, the spin sum rules
require them to be subdominant to the $\frac{3}{2}$ states. The most
natural solution for all the SV sum rules would be if the lowest 
$\frac{3}{2}$ states with $\epsilon_{3/2}\simeq 450\MeV$ have
$|\tau_{3/2}|^2\approx 0.3$, while for the $\frac{1}{2}$ states 
$|\tau_{1/2}|^2\approx 0.07\:\mbox{to}\:0.12$ with
$\epsilon_{3/2}\approx  300\:\mbox{to}\:500\MeV$.

Possible resolutions of this apparent contradiction have been
discussed. Strictly speaking, higher $P$-wave excitations can make up
for the wrong share between the contributions of the lowest states. This
possibility is not too appealing, however. In most known cases,
additionally, the lowest states in a given channel tend to saturate
the sum rules with a reasonable accuracy.

A certain loophole remains in that the experimental information comes
from the properties of the charmed mesons, which implies, generally
significant corrections. For instance, the classification itself over
the light cloud angular momentum $j$ relies on the heavy quark
limit. However, one probably needs a good physical reason to have 
the hierarchy between the finite-$m_c$ heirs of the $\frac{1}{2}$ and 
$\frac{3}{2}$ states inverted, rather than only 
reasonably modified compared to the heavy quark limit. 

Clearly, a too light actual
charm introduces significant practical complications here. Lattice
simulations can be of much help in this respect. There are ideas of
how to address this problem on the lattice in the most direct way. 

I think the clarification of this apparent contradiction between the
theoretical expectations and the existing measurements, together with
gaining better understanding of the saturation for both singlet and
spin sum rules, is an important task. It requires both new
theoretical insights and more detailed experimental data.
\vspace*{2mm}

\noindent
{\bf Conclusions}~ The dynamic QCD-based theory of inclusive 
heavy flavor decays has finally undergone critical experimental checks
sensitive to the nonperturbative contributions to 
the semileptonic $B$
decays. Experiment finds consistent values of the heavy quark
parameters extracted from quite different measurements once theory is
applied properly. The heavy quark parameters emerge close to the
theoretically expected values.
The perturbative corrections to the higher-dimension nonperturbative
heavy quark operators in the OPE have become the main limitation on
theory accuracy; this is likely to change in the foreseeable future. 

Inclusive decays can also provide important information for a number
of individual heavy flavor transitions.  The
$B\tto D\,\ell\nu$ decays may actually be accurately treated. The
successes in the dynamic theory of $B$ decays put new range of
problems in the focus; in particular, the issue of saturation of the
SV sum rules requires close scrutiny from both theory and experiment.

\section*{Acknowledgments}
I am grateful to D.~Benson, I.~Bigi, P.~Gambino, M.~Shifman,
A.~Vainshtein, to many colleagues from BaBar, CLEO and DELPHI, 
in particular O.~Buchmueller, V.~Luth and P.~Roudeau, for 
close collaboration and discussions.
This work was supported in part by the NSF under grant number
PHY-0087419.

\end{document}